\documentclass[journal]{IEEEtran}
\usepackage{cite}

\ifCLASSINFOpdf
\else
\fi
\usepackage[cmex10]{amsmath}
\usepackage{multirow}
\usepackage{graphicx}
\usepackage{array}
\usepackage{epstopdf}

\begin{document}
%
\title{A Unified Residential Energy Cost Optimization Model for Smart Grid - Significance and Challenge}
%
%
%

\author{Muhammad~Raisul~Alam\textsuperscript{1},
        Marc~St-Hilaire\textsuperscript{2},
        and~Thomas~Kunz\textsuperscript{3}
				
\textsuperscript{1,2,3}Department of Systems and Computer Engineering, Carleton University, Ottawa, ON, Canada 

\textsuperscript{2}School of Information Technology, Carleton University, Ottawa, ON, Canada

\textsuperscript{1}raisul@sce.carleton.ca, \textsuperscript{2}marc\_st\_hilaire@carleton.ca, \textsuperscript{3}tkunz@sce.carleton.ca

}

\maketitle

\enlargethispage*{10pt}
\begin{abstract}
This article addresses the residential energy cost optimization problem in smart grid. To date, most of the previous research only consider a partial aspect of the cost optimization problem. As a result, they fail to analyze scenarios when the interconnected components along with their properties have to be considered simultaneously. The proposed model combines these partial models into a single unified cost optimization model. Therefore, it is able to analyze scenarios which are closer to practical implementation. Furthermore, it is useful to analyze the behavior of a population (\textit{e.g.}, smart buildings, smart cities, etc.) and properties of the components for specific scenarios (\textit{e.g.}, the impact of aggregate storage capacity, etc.). It allows energy trading in microgrid which introduces a cost fairness problem. It ensures Pareto optimality among the households which guarantees that no household will be worse off to improve the cost of others. Results show that it can maintain the user preferences and can react to a demand response program by rescheduling the household loads and sources. Finally, the paper addresses the challenge of the computational complexity of the proposed model, showing that solution time increases exponentially with the problem size and proposes possible approaches to solve this.      
\end{abstract}


\ifCLASSOPTIONpeerreview
 \begin{center} \bfseries EDICS Category: 3-BBND \end{center}
 \fi
%
\IEEEpeerreviewmaketitle

\enlargethispage*{10pt}
\section{Introduction}
%
%
%
%
\IEEEPARstart{A}{} smart home in the smart grid offers an optimized energy management solution in collaboration with the utility and the neighbors. The widespread participation of the users plays an important role to achieve the optimal benefit from the smart grid. The users are mostly motivated by the cost saving potential of smart homes. A smart home has diverse appliances, devices and equipment with different energy consumption profiles. The user has his own preferences for the appliance operations that ensure  a preferred comfort level. A smart meter acquires price signals advertised by the utility. Instead of simply depending on the utility, a smart home can generate its own energy via solar panels, wind turbines, geothermal plants and other renewable energy sources. The renewables show stochastic energy generation profiles correlated to the weather. One or multiple storage devices accumulate cheap and excessive energy to support future energy requirements, avoiding the need to acquire expensive energy from the grid during periods of high prices (for example, at peak loads). However, the usage of such storage devices imposes a cost overhead because of self-discharging and efficiency loss. Any surplus energy can be traded via a microgrid which ensures a collective minimization of energy cost and potentially maximizes individual profits. 

Energy cost is primarily optimized by rescheduling the appliances when energy is cheap (energy sources could be grid, microgrid or renewables), provisioning appliance power, interrupting the appliance operation, scheduling storage usage, and energy trading. This research proposes a unified cost optimization model for smart homes that considers diverse energy sources, storage, and loads constrained by user preferences and the electrical properties of the participating components. The proposed cost minimization model combines  the features and components which are reported in the previous research in a unified single model and analyzes the behavior of the overall system. Most of the previous research considered a partial problem and therefore failed to analyze the scenarios when various of theses components interact among themselves.  

Energy trading in the microgrid introduces a cost fairness problem among the participating households. As the model aims to reduce the energy buying cost from the utility for the entire microgrid area, it failed to address the cost optimization for each individual household. It was found that the optimization model using only a single objective function (minimize the sum of the energy costs and the disutility costs for all households in the microgrid) sometimes increases the energy cost of a few households while reducing the cost of others. This unfair cost distribution problem made the model unattractive to the participating parties. To address this problem, the model has been extended to an optimization problem with multiple objective functions that considers individual household energy cost. In multi-objective optimization, we need to ensure Pareto optimality, \textit{i.e.}, no household should be worse off to improve the cost of some other households. To support energy trading in the microgrid, the model has been transformed from a simpler Mixed Integer Linear Programming (MILP) problem to a complex non-convex Mixed Integer NonLinear Programming (MINLP) problem which introduces a new challenge related to the complexity of the solution time. This paper addresses the issue and discusses the future direction to solve this problem.  

The objective of this research is to minimize the cost from the users' perspective. In other words, given a list of appliances that must be used within a given time horizon, we would like to know how all users optimally schedule their energy consumption and energy sources such that the total cost of all users is minimized. It is important to note that the goal is not about load optimization or balancing, which may be useful for the utility to reduce peaks.

The reminder of this article is organized as follows. Section II explores the cost saving strategies for smart homes discussed in the literature. Section III proposes a non-convex MINLP model that unifies the components for cost minimization in smart homes. Section IV solves the proposed cost optimization model and analyzes the results. Section V presents a numerical analysis of the time complexity of the proposed model. Finally, Section VI concludes the article addressing the contributions and significance of the outcomes. 
\enlargethispage*{10pt}
\section{Literature Review}
The previous research identified that optimal rescheduling of the household loads according to the energy price is one of the effective ways of cost minimization. Linear Programming (LP) is the most popular optimization method reported in the literature \cite{Mohsenian-Rad2010, Bapat2011, Angelis2013, Zhu2012, Conejo2010, Marwan2012, Hubert2011}. Other researchers also used Convex Programming \cite{Hovgaard2013, Tsui2012} and Game Theory \cite{Atzeni2013} to minimize energy cost. 

Based on the user preference, the appliances are primarily divided into two types: shiftable and non-shiftable \cite{Bapat2011}. The non-shiftable appliances cannot be delayed but may be operated at reduced power. Rescheduling is applicable to the shiftable devices which can be delayed to relatively lower energy cost time periods. The delay preference for this type of appliances depends on user comfort. The shiftable appliances may also be operated at reduced power. 

Based on the power consumption profile, an appliance can be interruptible or uninterruptible \cite{Mohsenian-Rad2010, Bapat2011, Tsui2012}. An interruptible device can be turned off in the middle of its operation and resumed later. An uninterruptable appliance must be allowed to continue its operation until the task is finished. Both types of appliances may require different levels of power during task execution based on internal operational states. These appliance-specific properties are generally enforced by constraints. 

Load scheduling and power shifting create discomfort to the user by delaying the task or reducing the task quality (by reducing power). A few methods considered this disutility \cite{Mohsenian-Rad2010,Bapat2011}. Some of the papers considered disutility created by temperature variation \cite{Kowahl2010}. Performing a survey on the participating users is a method to measure this disutility \cite{Georgievski2012}. A survey is more appropriate than other hypothetical assumptions because it is based on human evaluation. Another method to measure disutility is to apply a penalty whenever an appliance operation is delayed or operated at reduced power \cite{Mohsenian-Rad2010,Bapat2011}. The previous research introduced a disutility cost in addition to the energy cost. The efficiency of this type of model depends on the effective calculation of the disutility cost. The energy cost is a monetary value. Therefore, the disutility cost should be an equivalent monetary value. In economic theory, disutility is expressed using an approximation function. In the literature, the disutility is expressed as a function of power or delay or both. However, disutility varies according to user, appliance and situation. To adjust for this variation, different user-defined scaling parameters are used for different users and appliances \cite{Mohsenian-Rad2010}. The previous research rarely considered the situations when no feasible solution of the problem is achievable. De Angelis \textit{et al.} proposed a solution of this problem by discarding lower priority load \cite{Angelis2013}.

It was also reported that optimized utilization of storage devices (\textit{e.g.}, thermal, electrical) has an impact on energy cost. The space heating and hot water storage of the household can be considered as thermal storages. The battery of an Electric Vehicle (EV) is considered as an electrical storage when it is plugged in. The system looses energy during the energy storing process (\textit{e.g.}, charging, heating). Each type of storage shows a gradual spontaneous energy loss (\textit{e.g.}, self-discharging, temperature loss due to external environment). 

De Angelis \textit{et al.}, Zhang \textit{et al.}, and Hopkins \textit{et al.} modeled the energy storage using linear constraints and proposed LP models to optimize energy cost \cite{Angelis2013, Zhang2014, Hopkins2012}.  Arabali \textit{et al.} proposed a Genetic Algorithm (GA) based optimization method to select the optimal storage capacity for the smart grid \cite{Arabali2013}. Vytelingum \textit{et al.} proposed a game-theoretic framework to analyze the effect of household storage on the energy price \cite{Vytelingum2010}. 

Energy storage may have an unintended impact on the smart grid. It is evident that an unplanned installation of storage at the user side may create a new critical peak energy demand. Sometimes, rather than saving energy cost, storage systems may increase energy cost \cite{Vytelingum2010}. In contrast, some research has reported that a distributed storage may be more beneficial than distributed generators \cite{Poonpun2008}. 

It is evident from the literature that the usage of renewable sources and advanced planning of renewable energy utilization based on the predicted generation quantity reduces energy cost. Arabali \textit{et al.} proposed a probabilistic GA based optimization using the 2 Point Estimation method to minimize PhotoVoltaics and wind generation installation cost and increase energy usage efficiency \cite{Arabali2013}. Bilil \textit{et al.} proposed a multi-objective optimization model to optimize annualized cost when diverse renewable energy sources are used for energy generation \cite{Bilil2014}. De Angelis \textit{et al.} addressed the significance of renewable energy sources for cost optimization in the smart grid \cite{Angelis2013}. The paper used a MILP model to optimize energy generation considering the given load. Zhang \textit{et al.} considered multiple homes in a smart building which share common distributed energy resources such as Combined Heat and Power (CHP) generators and boilers and used a MILP model to optimized cost \cite{Zhang2014}. Hopkins \textit{et al.} proposed a LP model to analyze the impact of  distributed generations on cost optimization \cite{Hopkins2012}. Bilil \textit{et al.} showed that a combination of different energy sources may result in a more predictable energy generation \cite{Bilil2014}. The current research trends have not yet identified the most effective combination of different energy sources.

Previous studies claimed that energy trading among the households by forming a microgrid plays an important factor for cost minimization in the smart grid. Ramachandran \textit{et al.} proposed a risk-based auction strategy to maximize profit while selling energy to the energy market \cite{Ramachandran2011}. Wang \textit{et al.} developed a Particle Swarm Optimization based negotiating agent to facilitate energy trading in the smart grid \cite{Wang2012}. Vytelingum \textit{et al.} developed a trading agent based on the Continuous Double Auction strategy \cite{Vytelingum2010}. Ilic \textit{el al.} described an energy market, named NOBEL \cite{NOBEL} to evaluate market driven demand response of electricity trading \cite{Ilic2012}. 

Based on these reviewed papers, it is apparent that different types of auctions are the most popular strategies for energy trading in the energy market. The auction strategies could be improved by using algorithms that predict the opponents' behavior \cite{Wang2012}. These algorithms use functions which adaptively determine the risk attitude of the users based on the previous trading history \cite{Wang2012}. The risk attitude of the users could be classified into risk seeking (high profit but higher risk of untraded energy), risk averse (low profit and lower risk of untraded energy) and risk neutral attitudes \cite{Ramachandran2011}. The performance of the intelligent algorithms is compared with the Zero Intelligence \cite{Gode1993} auction strategy which is based on random bid and ask prices  \cite{Vytelingum2010, Ilic2012}.

In summery, most of the existing research failed to address a comprehensive discussion of smart homes, considering all important potential avenues for cost optimization. Our work combines these potential strategies into a single unified cost optimization model and analyzes the behavior of the overall system. 
\enlargethispage*{10pt}
\section{The Proposed Unified Model}
A non-convex MINLP model has been developed to optimize the energy cost for the proposed problem. The following notation is used to describe input parameters and decision variables.
\enlargethispage*{10pt}
\subsection{Input Parameters}
\subsubsection{Sets}
\begin{itemize}
\item$\mathit{H}$, set of timeslots representing the scheduling horizon  where $\mathit{h \in H}$ is the $\mathit{h}$-th timeslot of set $\mathit{H}$. 
\item$\mathit{I}$, set of appliances where $\mathit{i \in I}$ is the $\mathit{i}$-th appliance of set \mbox{$\mathit{I}$}.   
\item$\mathit{K}$, set of households and $\mathit{k \in K}$ is the \mbox{$\mathit{k}$-th} household of set $\mathit{K}$. 	
\item$\mathit{U}$, set of uninterruptible appliances where $\mathit{U\subset I}$.
\end{itemize}
\subsubsection{Constants}
\begin{itemize}
\item$\mathit{\beta_{k,i}}$, maximum allowable delay of the $\mathit{i}$-th appliance of the \mbox{$\mathit{k}$-th} household. 
\item$\mathit{d_{k,i}}$, disutility factor of the $\mathit{i}$-th appliance of the \mbox{$\mathit{k}$-th} household. The disutility related to every appliance is not the same. We can control the delay preference of each appliance by controlling the value of $\mathit{d_{k,i}}$. 
\item$\mathit{E_k}$, storage efficiency of the \mbox{$\mathit{k}$-th} household.
\item$\mathit{GP_h}$, energy price of the utility grid at timeslot $\mathit{h}$. It is assumed that the electricity price changes at the starting second of a timeslot and stays at the same price until the end of that timeslot. 
\item$\mathit{IE_k}$, initial storage energy of the \mbox{$\mathit{k}$-th} household.
\item$\mathit{L_k^{max}}$, maximum grid power limit of the \mbox{$\mathit{k}$-th} household imposed by the utility or by the electric fuse (per scheduling timeslot).
\item$\mathit{MaxC_k}$, maximum storage capacity of the \mbox{$\mathit{k}$-th} household.
\item$\mathit{MinC_k}$, minimum storage capacity of the \mbox{$\mathit{k}$-th} household. If the storage energy goes below the minimum level, it reduces the lifespan of the storage.
\item$\mathit{N}$, number of timeslots with fixed duration.  
\item$\mathit{p_{k,i}}$, power consumption of the $\mathit{i}$-th appliance of the \mbox{$\mathit{k}$-th} household.
\item $\mathit{r_{k,i,h}}$,  reservation time of an appliance which represents the time when the scheduler gets a request to start a specific appliance. $\mathit{r_{k,i,h}=1}$ means that operation of appliance $\mathit{i}$ of the \mbox{$\mathit{k}$-th} household is requested in timeslot $\mathit{h}$. Alternatively,  $\mathit{r_{k,i,h}=0}$ means that appliance $\mathit{i}$ of the \mbox{$\mathit{k}$-th}household is not requested in timeslot $\mathit{h}$. If an appliance gets multiple requests in the same time horizon, it is considered as multiple (virtual) appliances. In this case, the scheduler optimizes the execution time as if it had multiple similar loads and there should be a constraint to ensure that the end time of the previous instance is at or before the start time of the later instance.  
\item$\mathit{RQ_{k,h}}$, amount of generated renewable energy of the \mbox{$\mathit{k}$-th} household at timeslot $\mathit{h}$.
\item$\mathit{SD_k}$, self-discharging coefficient of the storage of the \mbox{$\mathit{k}$-th} household.
\item$\mathit{SP_k}$, power required to charge the storage of the \mbox{$\mathit{k}$-th} household.
\item$\mathit{t_{k,i}}$, duration of the running time of the $\mathit{i}$-th appliance of the \mbox{$\mathit{k}$-th} household. 
\end{itemize}
\enlargethispage*{10pt}
\subsection{Decision Variables}
\begin{itemize}
\item$\mathit{BE_{k,h}}$ is a positive real variable which represents the energy used from the storage by the \mbox{$\mathit{k}$-th} household at timeslot $\mathit{h}$. 
\item$\mathit{GE_{k,h}}$ is a positive real variable which represents the energy drawn from the utility grid by the \mbox{$\mathit{k}$-th} household at timeslot $\mathit{h}$.  
\item$\mathit{IC_{k,h}}$ is a Boolean vector which represents whether the storage is in charging stage or not. $\mathit{IC_{k,h}=1}$ means the storage of household $\mathit{k }$ is in charging stage at timeslot $\mathit{h}$.
\item$\mathit{ME_{k,h}}$  is the energy traded with the microgrid by the \mbox{$\mathit{k}$-th} household at timeslot $\mathit{h}$. A positive value of $\mathit{M_{k,h}}$ means the \mbox{$\mathit{k}$-th} household is a buyer at timeslot $\mathit{h}$. A negative value means the household is a seller.   
\item$\mathit{MP_h}$ is a positive real variable which represents the price of microgrid energy at timeslot $\mathit{h}$.
\item$\mathit{MQ_{k,h}}$ is the demand (or supply) of microgrid energy of the \mbox{$\mathit{k}$-th} household at timeslot $\mathit{h}$. A positive value represents the minimum energy demand of the household. A negative value represents the maximum amount of energy the household can sell to the microgrid. 
\item$\mathit{RE_{k,h}}$ is a positive real variable which represents the energy used from the renewable sources by the \mbox{$\mathit{k}$-th} household at timeslot $\mathit{h}$.  
\item$\mathit{SE_{k,h}}$ is a positive real variable which represents the stored energy in the storage of the \mbox{$\mathit{k}$-th} household at timeslot $\mathit{h}$.  
\item$\mathit{S_{k,i,h}}$ is a Boolean vector which represents the execution time of the appliances. $\mathit{S_{k,i,h} =1}$ means that appliance $\mathit{i}$ runs at timeslot $\mathit{h}$ in household $\mathit{k}$.
\item$\mathit{US_{k,i,h}}$ is a Boolean vector which represents the start time of the uninterruptible appliances. $\mathit{US_{k,i,h} =1}$ means that the uninterruptible appliance $\mathit{i}$ starts at timeslot $\mathit{h}$ in household $\mathit{k}$.
\item$\mathit{\tau_{k,i}}$ is the end time of the task executed by appliance $\mathit{i}$ at $\mathit{k}$-th household. $\mathit{\tau_{k,i}}$ is a positive integer variable that can take values in the following range: $\mathit{[1, N]}$.
\end{itemize}
\enlargethispage*{10pt}
\subsection{Energy Function}
The energy cost of the $\mathit{k}$-th household, denoted by $\mathit{CE_{k}}$, is a function of utility price, the power drawn from the grid, microgrid price and energy traded with the microgrid. This can be expressed in a cost function which is given in (\ref{eqn_1}).
\begin{equation}
\label{eqn_1}
CE_{k}=\sum\limits_{h\in H}^{} GP_h \cdot GE_{k,h} +\sum\limits_{h\in H}^{} MP_h \cdot ME_{k,h}
\end{equation}

To optimize the electricity cost in a day, the appliances may be scheduled in timeslots where electricity prices are relatively low. Therefore, to get the optimal savings, $\mathit{CE_{k}}$  should be minimized.
\enlargethispage*{10pt}
\subsection{Disutility Function}
If the scheduler tries to minimize $\mathit{CE_{k}}$, it may create discomfort to the user because it may cause the appliances to be delayed for a long period of time. To consider this discomfort while trying to reduce costs, a disutility function is used to compensate for the inconvenience created by delaying an appliance operation. The disutility function, denoted by $\mathit{CD_{k}}$ for the $\mathit{k}$-th household, is expressed in (\ref{eqn_2}).
\begin{equation}
\label{eqn_2}
CD_{k} = \sum\limits_{i\in I} d_{k,i} \Bigg( \tau_{k,i} - \bigg(\sum\limits_{h\in H} r_{k,i,h}\cdot h + t_{k,i}-1\bigg)\Bigg)
\end{equation}

Here, $\mathit{\sum\nolimits_{h \in H} r_{k,i,h} \cdot h}$ is the reservation timeslot which is calculated from the reservation times. An appliance is reserved only once in the time horizon (which means $\mathit{\sum\nolimits_{h \in H} r_{k,i,h}=1}$). Therefore, the multiplication of $\mathit{r_{k,i,h}}$ by the corresponding timeslot will provide the reservation time. Hence, $\mathit{\sum\nolimits_{h\in H} r_{k,i,h} \cdot h+ t_{k,i} -1}$ represents the end time of the appliance, if it has been started immediately after reservation. Therefore, if we subtract this end time from the scheduled end time $\mathit{\tau_{k,i}}$, we obtain the delay from the reservation time. The disutility factor $\mathit{d_{k,i}}$ is an adjustable coefficient which is defined according to the users' tolerance of delay per appliance. The higher the value of $\mathit{d_{k,i}}$, the more disutility such a delay will generate and to compensate this, the scheduler will reduce the delay.
\subsection{The Optimization Model}
The total cost of the 1st household is denoted by $\mathit{C_1}$, the total cost of the 2nd household is denoted by $\mathit{C_2}$ and so on. In general, the total cost of the \mbox{$\mathit{k}$-th} household is,
\begin{equation}
\label{eqn_3}
C_k=CE_{k}+CD_{k}
\end{equation}
The objective is to minimize all cost functions as follows, 
 \begin{equation}
\label{eqn_4}
min (C_1,C_2, \dots, C_k)
\end{equation}
Objective (\ref{eqn_4}) is a multi-objective MILP optimization problem which is implemented using a single objective function that minimizes the sum of the energy costs and disutility cost for all households in the microgrid (Constraint (\ref{eqn_23}) discussed later). The optimal solution should satisfy the following constraints. 

\subsubsection{Energy Conservation Constraints}
For a specific household, the total energy consumed by all appliances and the storage in a specific timeslot should be the same as the total energy supplied by the grid, renewable sources, microgrid and storage in that timeslot. This relation has been expressed in (\ref{eqn_5}).
\setlength{\arraycolsep}{0.0em}
\begin{eqnarray}
\label{eqn_5}
\sum\limits_{i\in I}^{}S_{k,i,h} \cdot p_{k,i} + IC_{k,h} \cdot SP_k = GE_{k,h} + BE_{k,h} \nonumber\\
+ RE_{k,h} + ME_{k,h} , (k \in K, h\in H)
\end{eqnarray}
\setlength{\arraycolsep}{5pt}If the microgrid energy $\mathit{ME_{k,h}}$ is negative, it becomes a load which means the $\mathit{k}$-th household sells energy to the microgrid during the $\mathit{h}$-th timeslots. If it is positive, it is an energy source \textit{i.e.}, the $\mathit{k}$-th household buys energy from the microgrid during the $\mathit{h}$-th timeslots. For this constraint, the microgrid can never be both a source and a load at the same time.
\subsubsection{Stored Energy Constraints}
In the first timeslot, storage energy is measured using (\ref{eqn_6}) which is a function of initial energy, charging and discharging at the 1st timeslot, efficiency loss and self-discharging loss.\setlength{\arraycolsep}{0.0em}
\begin{eqnarray}
\label{eqn_6}
SE_{k,1}=IE_k\cdot SD_k + IC_{k,1} \cdot SP_k \cdot E_k - BE_{k,1}, \nonumber\\(k \in K)
\end{eqnarray}
\setlength{\arraycolsep}{5pt}The stored energy in the subsequent timeslots depends on the energy during the immediate previous timeslot, the energy stored in the current timeslot considering the efficiency loss, and the self-discharging loss, which has been expressed in (\ref{eqn_7}). Self-discharging only applies to the stored energy from the immediate previous timeslot. 
\setlength{\arraycolsep}{0.0em}
\begin{eqnarray}
\label{eqn_7}
SE_{k,h}=IE_{k,h-1}\cdot SD_k + IC_{k,h} \cdot SP_k \cdot E_k - BE_{k,h}, \nonumber\\(k \in K, {h\in H:h\neq 1})
\end{eqnarray}
\setlength{\arraycolsep}{5pt}\subsubsection{Storage Capacity Constraints}
A storage must have sufficient energy to act as a source. It should not exceed the maximum storage capacity and should not go below the minimum energy level. Inequalities (\ref{eqn_8}) and (\ref{eqn_9}) limit the maximum and minimum stored energy of the storage respectively.
\setlength{\arraycolsep}{0.0em}
\begin{eqnarray}
\label{eqn_8}
SE_{k,h} \le MaxC_k, (k \in K, h\in H)\\
\label{eqn_9}
SE_{k,h} \ge MinC_k, (k \in K, h\in H)
\end{eqnarray}
\setlength{\arraycolsep}{5pt}\subsubsection{Task Duration Constraints}
Constraint (\ref{eqn_10}) maintains the total duration of a task. It ensures that the sum of all the elements of the scheduling vector of an appliance equals to the duration of that appliance. 
\setlength{\arraycolsep}{0.0em}
\begin{eqnarray}
\label{eqn_10}
\sum\limits_{h\in H}^{} S_{k,i,h} = t_{k,i}, (k \in K, h\in H)
\end{eqnarray}
\setlength{\arraycolsep}{5pt}\subsubsection{Renewable Energy Availability Constraints}
The total energy used from the renewable sources should be less than or equal to the total generated energy by the renewables as expressed in (\ref{eqn_11}).
\setlength{\arraycolsep}{0.0em}
\begin{eqnarray}
\label{eqn_11}
RE_{k,h} \le RQ_{k,h}, (k \in K, h\in H)
\end{eqnarray}
\setlength{\arraycolsep}{5pt}\subsubsection{Energy Conservation Constraints for the Microgrid}
 Constraint (\ref{eqn_12}) imposes that the total energy sold in the microgrid must be equal to the total energy bought from the microgrid.  
\setlength{\arraycolsep}{0.0em}
\begin{eqnarray}
\label{eqn_12}
\sum\limits_{k \in K}^{} ME_{k,h} = 0, (h\in H)
\end{eqnarray}
\setlength{\arraycolsep}{5pt}\subsubsection{Microgrid Energy Price Constraints}
The energy price of the microgrid in a specific timeslot will be greater than or equal to 0 and cannot be greater than the grid energy price ((\ref{eqn_13}) and (\ref{eqn_14})). 
\setlength{\arraycolsep}{0.0em}
\begin{eqnarray}
\label{eqn_13}
MP_h\ge 0, (h\in H)\\
\label{eqn_14}
MP_h\le GP_h, (h\in H)
\end{eqnarray}
\setlength{\arraycolsep}{5pt}If the microgrid energy price exceeds the grid price, the household should buy energy from the grid.
\subsubsection{Energy Constraints while Trading in the Microgrid}
The energy surplus or demand is the difference between the total energy consumption and generation in a specific timeslot.  (\ref{eqn_15}) calculates the amount of energy that is available to trade in the microgrid.  
\begin{eqnarray}
\label{eqn_15}
MQ_{k,h}=\sum\limits_{i\in I}^{} S_{k,i,h} \cdot p_{k,i} + IC_{k,h}\cdot SP_k - GE_{k,h}\nonumber\\ - RQ_{k,h} - SE_{k,h} - BE_{k,h} + MinC_k, (k \in K, h\in H)
\end{eqnarray}
If the user is a buyer, Constraint (\ref{eqn_16}) defines the minimum energy required by the household from the microgrid. If the user is a seller, this constraint limits the maximum amount of energy that a household can sell to the microgrid. 
\setlength{\arraycolsep}{0.0em}
\begin{eqnarray}
\label{eqn_16}
M_{k,h} \ge MQ_{k,h}, (k \in K, h\in H)
\end{eqnarray}
\setlength{\arraycolsep}{5pt}\subsubsection{Reservation Time Constraints}
Constraint (\ref{eqn_17}) specifies that all executions must start after (or at) the reservation time. Without this constraint, an appliance could be scheduled to run before it has been requested. Here, $\mathit{\sum\nolimits_{h\in H} r_{i,h}\cdot h}$ refers to the reservation timeslot. 
\setlength{\arraycolsep}{0.0em}
\begin{eqnarray}
\label{eqn_17}
\sum\limits_{h\in H}^{} S_{k,i,h} = \sum\limits_{h=\sum\nolimits_{h\in H} r_{k,i,h}\cdot h}^N S_{k,i,h}, \nonumber\\  (k \in K, i\in I)
\end{eqnarray}
\setlength{\arraycolsep}{5pt}\subsubsection{Relationship between the Scheduling Vector and the End Time Constraints}
Inequality (\ref{eqn_18}) binds $\mathit{S_{k,i,h}}$  with $\mathit{\tau_{k,i}}$. These two decision variables depend on each other: for any appliances, the last execution time should also be the end time. It also implicitly expresses that, for the earliest possible end time, execution time cannot be zero.
\setlength{\arraycolsep}{0.0em}
\begin{eqnarray}
\label{eqn_18}
S_{k,i,h}\cdot h \le \tau_{k,i}, (k \in K, i \in I,  h\in H)
\end{eqnarray}
\setlength{\arraycolsep}{5pt}\subsubsection{Maximum Execution End Time Limit Constraints}
Constraints (\ref{eqn_19}) imposes that the end time must be before (or at) the user defined maximum execution time limit.
\setlength{\arraycolsep}{0.0em}
\begin{eqnarray}
\label{eqn_19}
\tau_{k,i}\le \beta_{k,i}, (k \in K, i \in I)
\end{eqnarray}
\setlength{\arraycolsep}{5pt}\subsubsection{Uninterruptibility Constraints}
Constraints (\ref{eqn_20}) and (\ref{eqn_21}) define that if an uninterruptable appliance starts running, it will keep running until it completes its operation. Without these constraints, an uninterruptible appliance may be interrupted.
\setlength{\arraycolsep}{0.0em}
\begin{eqnarray}
\label{eqn_20}
\sum\limits_{d=0}^{t_{k,i}-1} S_{k,i,{h+d}} - t_{k,i} \ge -t_{k,i}(1-US_{k,i,h}),\nonumber\\
(k \in K, i\in U, h=[1, N-t_{k,i}+1]) \\
\label{eqn_21}
\sum\limits_{h=1}^{N-t_{k,i}+1} US_{k,i,h}=1, (k \in K, i\in U)
\end{eqnarray}
\setlength{\arraycolsep}{5pt}\subsubsection{Utility Grid Max Power Limit Constraints}
Inequality (\ref{eqn_22}) limits the load of the utility grid per timeslot to a maximum power limit. Without this constraint, the appliances may be scheduled in such a way that they exceed the maximum grid power limit per household.  
\setlength{\arraycolsep}{0.0em}
\begin{eqnarray}
\label{eqn_22}
GE_{k,h} \le L_k^{max}, (k \in K, h\in H)
\end{eqnarray}
\setlength{\arraycolsep}{5pt}\subsection{Extended Model with Pareto Optimality}
The objective function specified in (\ref{eqn_4}) is a unified optimization problem that tries to minimize the sum of the energy cost and the disutility cost of all household in the microgrid. Sometimes an optimal solution will increase the energy cost of a few households to reduce the cost of others. This situation occurs when a household in essence buys more energy than it needs and gives the energy away for cheap to other households. Therefore, the model should ensure Pareto optimality to address the cost fairness problem.  We extended the objective function proposed in (\ref{eqn_4}) by adding a new function Y defined as the total cost of all households in that microgrid area, which is expressed in (\ref{eqn_23}).
\setlength{\arraycolsep}{0.0em}
\begin{eqnarray}
\label{eqn_23}
Y=\sum\limits_{k \in K}^{}\sum\limits_{h\in H}^{} GP_h\cdot GE_{k,h} + \sum\limits_{k \in K}^{}\nonumber\\ \sum\limits_{i\in I}^{} d_{k,i} \Bigg(\tau_{k,i}- \bigg(\sum\limits_{h\in H}^{} r_{k,i,h}\cdot h + t_{k,i}-1\bigg)\Bigg)
\end{eqnarray}
\setlength{\arraycolsep}{5pt}
Eq. (\ref{eqn_23}) does not require the cost of energy drawn/supplied from/to the microgrid because in a specific area, the total buying cost from the microgrid equals the total selling profit by all households. Therefore, the microgrid energy price does not have an impact on the total cost. Hence, $\mathit{Y}$ is a linear function that represents the total energy cost drawn from the utility grid by all households and the total disutility because of delayed tasks. 
The extended multi-objective optimization model is,
\setlength{\arraycolsep}{0.0em}
\begin{eqnarray}
\label{eqn_24}
min (Y, C_1, C_2, \dots , C_k)
\end{eqnarray}
\setlength{\arraycolsep}{5pt}subject to Constraints (\ref{eqn_5}) - (\ref{eqn_22}).
The extended model is solved by using $\mathit{Y}$ as the sole objective function and the remaining objective functions, $\mathit{C_k}$, are added as inequality constraints.  Suppose, for the $\mathit{k}$-th household, the minimum energy cost in the absence of the microgrid is represented by $\mathit{C_k^{NoTrade}}$. Therefore, the proposed multi-objective optimization problem is defined as:
\begin{equation}
\label{eqn_25}
min {~}Y
\end{equation}subject to:
\begin{equation}
\label{eqn_26}
C_k \le C_k^{NoTrade}, (k \in K)
\end{equation}and all other Constraints (\ref{eqn_5}) - (\ref{eqn_22}). 

The following analysis provides significant information about the model which helps to implement the algorithm practically. The energy cost of a household, $\mathit{C_k^{NoTrade}}$, can be calculated when the household is considered as an isolated household, meaning that the household does not take part in the microgrid trading. Therefore, if we minimize (\ref{eqn_23}) with respect to (\ref{eqn_5}) - (\ref{eqn_11}) and (\ref{eqn_17}) - (\ref{eqn_22}), the solution will provide us with the maximum energy cost boundary of the specific household when it participates in microgrid. In (\ref{eqn_5}), microgrid energy, $\mathit{ME_{k,h}}$, is set to 0. This optimization excludes the microgrid-related constraints because it is calculating the cost of an individual household. 

Objective Function (\ref{eqn_25}) with constraints (\ref{eqn_5}) - (\ref{eqn_22}) and (\ref{eqn_26}) provides a Pareto-optimal solution of the problem. Due to constraint (\ref{eqn_26}), no household is paying more than what it used to pay before participating in the microgrid. Objective Function (\ref{eqn_25}) ensures that the overall cost in that microgrid neighborhood is minimized without causing any of the participating parties to be worse off. 

Obviously, there are numerous Pareto-optimal solutions that satisfy the above condition. It is possible to generate all Pareto-optimal points by iteratively decreasing the value of $\mathit{C_k^{NoTrade}}$. Abounacer \textit{et al.} proposed a solution approach using the epsilon constraint method that iteratively generates all Pareto-optimal solutions\cite{Abounacer2014}. However, it is computationally expensive because the solution time increases exponentially according to the number of households. Zhang \textit{et al.} proposed a lexicographic minimax method that distributes the cost savings in a proportional fair way among the households. However, it also results in a more complex optimization problem, requiring to solve the problem iteratively in polynomial time \cite{Zhang2014}. 

\section{Case Studies}
This section presents numerical analysis of the unified model described in Section III. The proposed model has been implemented in AMPL (A Mathematical Programming Language) \cite{Fourer2002,AMPL}, an algebraic modeling language for describing large-scale complex mathematical optimization problems. We used the Couenne (Convex Over and Under ENvelopes for Nonlinear Estimation) solver \cite{Belotti2009} available as a part of the NEOS (Network-Enabled Optimization System) Server \cite{NEOS}. NEOS is a free Internet-based service for solving optimization problems. The Couenne solver aims at finding the global optima of non-convex MINLPs. It implements linearization, bound reduction, and branching methods within a spatial branch-and-bound framework. We used Couenne 0.4.3 for all the results presented in this article. 

The case studies consider simple but complete scenarios to demonstrate the impact of the different components of the model. The scenarios consider 8 timeslots and 2 households. Each household has 2 appliances. The first appliance (App1) is interruptible and the second appliance (App2) is uninterruptible. If it is not mentioned explicitly elsewhere, the appliance characteristics are those shown in Table \ref{table_app_char}. The appliances have different duration and power requirements (per timeslot), and all should be scheduled between the first and the last scheduling timeslots, such that they complete by the 8th scheduling timeslot. The maximum power per household that could be drawn from the grid is limited to 20 units.

\begin{table}[!t]
\renewcommand{\arraystretch}{1.3}
\caption{Appliance Characteristics}
\label{table_app_char}
\centering
\begin{tabular}{l>{\raggedright}p{.7cm}>{\centering}p{0.7cm}>{\centering}p{.7cm}>{\centering}p{.7cm}>{\centering}p{.7cm}>{\centering}p{.7cm}}
\hline
Household & Appli-ance	& Dura-tion (Time-slots)	& Power (Unit)	& Disuti-lity Factor	& {Reser-vation Time-slot} &	Max Delay \tabularnewline
\hline
\multirow{2}{*} {Household 1} & App1	& 5	&1	 &0.01	&1	 &8\tabularnewline
 &App2   	& 2	&4	 &0.01	&1	 &8\tabularnewline

\multirow{2}{*} {Household 2} & App1	& 3	&3 &0.01	&1	 &8\tabularnewline
 &App2  	& 4	&2	 &0.01	&1	 &8\tabularnewline
\hline
\end{tabular}
\end{table}

\begin{table}[!t]
\renewcommand{\arraystretch}{1.3}
\caption{Storage Characteristics}
\label{table_sto_char}
\centering
\begin{tabular}{l>{\centering}p{1.5cm}>{\centering}p{1.5cm}}
\hline
Characteristics & Household  1	&Household 2 \tabularnewline
\hline
Initial Energy (Unit)	& 3	& 5\tabularnewline
Power (Unit)	& 1 &	2\tabularnewline
Max Capacity (Unit)	& 5	& 5\tabularnewline
Min Capacity (Unit)	 & 3	& 3\tabularnewline
Efficiency	& 80\%	 & 90\%\tabularnewline
Self-Discharging Rate \tabularnewline(Per Timeslots)	& 0.01\%	& 0.01\%\tabularnewline
\hline
\end{tabular}
\end{table}

Each household also has a storage device with characteristics outlined in Table \ref{table_sto_char}. The storage devices have a maximum capacity of 5, should never drop below 3, have different initial charges, different charging efficiencies, and discharge at a rate of 1\% per timeslot. The device in household 1 can be charged in increments of 1, while the device in household 2 can only be charged in increments of 2 units per timeslot.
\enlargethispage*{10pt}
\subsection{User Preference}
The user uses disutility factors to set his delay preferences for the appliances. The disutility factor applies an inconvenience cost if the appliance operation is delayed. The higher the value of the disutility factor, the less the model will tolerate appliance execution delay. Tables \ref{table_dis_1} and \ref{table_dis_2} outline the energy consumption profiles and how the appliances are scheduled, given a specific grid energy price profile (listed at the top) for household 1. In this example, energy from the grid is relatively cheap early and late in the scheduling period, but very expensive during timeslots 4-7. The next two blocks in the tables specify what energy is available and how much energy is actually drawn upon from various sources. The energy available from the grid is capped at 20 units to model a per-household max load on the power grid. The available storage energy is determined initially by the initial charge and then evolves over time to reflect charging/discharging and self-discharging. The energy profile for the renewable devices is assumed to be known/given, for example by using historical data and Markov chain models to predict it, see for example \cite{Ardakanian2011}. The fourth block in the table specifies how much of that available energy is actually drawn upon by indicating which appliance is scheduled to run in that timeslot and whether we charge the storage device. For example, at timeslot 8 in Table \ref{table_dis_1}, both appliances are scheduled to operate, requiring a total of 5 units of energy. Based on the block right above, all 5 units are drawn from the grid, as the grid price in that timeslot is low. In this example, the optimizer determines that it is more efficient to charge the storage with 1 unit of cheap grid energy in timeslot 1 (and with free energy from one of the renewable in timeslots 4 and 6) and draw on this in timeslot 7. Appliance 1 (which is interruptible) is scheduled to operate in timeslots 1, 4, 5, 6, and 8, for a total duration of 5 timeslots. Appliance 2 (which is non-interruptible) is scheduled to operate in timeslots 7 and 8. It may appear surprising that Appliance 1 is scheduled to operate during timeslots with relatively expensive grid energy prices (timeslots 4 to 6), but at these times the appliance is powered from the energy delivered by the renewables, which have a cost of 0. The total energy cost to the household is 6.99 which is the sum of the total energy cost (6.9) and the total disutility cost (0.09) of all appliances. 

\begin{table}[!t]
\renewcommand{\arraystretch}{1.3}
\caption{Household 1 Cost without Microgrid Trading and Low Disutility Factors (Disutility Factors=0.01)}
\label{table_dis_1}
\centering
\begin{tabular}{>{\raggedright}p{.9cm} >{\raggedright}p{1.2cm} >{\centering}p{.4cm} >{\centering}p{0.4cm} >{\centering}p{0.3cm} >{\centering}p{0.3cm} >{\centering}p{0.3cm} >{\centering}p{0.3cm} >{\centering}p{0.3cm} >{\centering}p{0.3cm} }
\hline
Timeslot & {} & 1 & 2 & 3 & 4 & 5 & 6 & 7 & 8 \tabularnewline
\hline
Price	&Grid	&0.7&	1	&1.2	&1.5	&2	&1.7&	1.5&	0.5\tabularnewline\hline
{Energy } & Grid	&20&20&	20&	20&	20&	20	&20	&20\tabularnewline
{Availa-}											 		& Storage	&3.59	&3.56&	3.52&	4.29&	4.24	&5	&3.03	&3\tabularnewline
{bility}													& Renewables	&0	&0	&0&	2&	1	&2&	0	&0\tabularnewline
\hline
{Energy} & Grid	&1.82	&0&	0	&0	&0	&0	&2.08	&5\tabularnewline
{Source}											 		& Storage	&0.18&	0&	0	&0	&0	&0	&1.92	&0\tabularnewline
{}													& Renewables	&0	&0	&0	&2	&1	&2&	0	&0\tabularnewline
\hline
{Load} & Storage Charging	&1&	0&	0	&1&	0&	1	&0	&0 \tabularnewline
{}											 		& App1	&1	&0	&0	&1&	1&	1	&0&	1\tabularnewline
{}													& App2	&0	&0	&0	&0	&0&0&	1	&1\tabularnewline
\hline
\multicolumn{10}{c}{Energy Cost = 6.9, Disutility Cost = 0.09, and Total Cost = 6.99} \tabularnewline
\hline

\end{tabular}
\end{table}

\begin{table}[!t]
\renewcommand{\arraystretch}{1.3}
\caption{Household 1 Cost without Microgrid Trading and High Disutility Factors (Disutility Factors = 5)}
\label{table_dis_2}
\centering
\begin{tabular}{>{\raggedright}p{.9cm} >{\raggedright}p{1.2cm} >{\centering}p{.4cm} >{\centering}p{0.4cm} >{\centering}p{0.3cm} >{\centering}p{0.3cm} >{\centering}p{0.3cm} >{\centering}p{0.3cm} >{\centering}p{0.3cm} >{\centering}p{0.3cm} }
\hline
Timeslot & {} & 1 & 2 & 3 & 4 & 5 & 6 & 7 & 8 \tabularnewline
\hline
Price	&Grid	&0.7&	1	&1.2	&1.5	&2	&1.7&	1.5&	0.5\tabularnewline
\hline
{Energy } & Grid	&20&20&	20&	20&	20&	20	&20	&20\tabularnewline
{Availa-}				& Storage	&3.77	&	3.73		&3		&3.77	&	3.73		&3.69		&3.66	&	3.62\tabularnewline
{bility}													& Renewables	&0	&	0		&0		&2	&	1		&2		&0	&	0\tabularnewline
\hline
{Energy} & Grid	&6		& 5		&0.31	&	0		&0		&0	&	0	&	 0\tabularnewline
{Source}							& Storage	&0	&	0	&	0.69		&0		&0	&	0		&0	&	0\tabularnewline
{}													& Renewables	&0	&	0	&	0	&	2		&1		&0	&	0		&0\tabularnewline
\hline
{Load} & Storage Charging	&1	&	0	&	0		&1	&	0	&	0		&0		&0\tabularnewline
{}											 		& App1	&1		&1		&1		&1	&	1	&	0	&	0		&0\tabularnewline
{}													& App2	&1	&	1		&0		&0		&0		&0		&0	&	0\tabularnewline
\hline
\multicolumn{10}{c}{Energy Cost = 9.57, Disutility Cost = 0, and Total Cost = 9.57} \tabularnewline
\hline

\end{tabular}
\end{table}

Table \ref{table_dis_2} likewise shows the available energy, consumed energy, and its use, for household 1 with higher disutility factors. The only difference between Table \ref{table_dis_1} and Table \ref{table_dis_2} is that for all appliances, the disutility factors used in Table \ref{table_dis_1} are very low (set to 0.01) and the disutility factors used in Table \ref{table_dis_2} are high (set to 5). The total cost to the household is now 9.57 which is only composed of the energy cost. Due to the higher disutility factors, the optimizer did not delay any appliance operation and hence the total disutility cost is 0. The table shows that the total cost is increased. This is because to comply with the user delay preference for the appliances, the optimizer buys expensive energy from the grid at timeslot 2 and 3 instead of timeslot 8 when grid energy is cheap. The optimizer did not use the free renewable energy at timeslot 6. At the end of the scheduling horizon in timeslot 8, the storage has unused energy because the optimizer had to charge the storage in timeslot 4 to maintain its minimum energy level but there was no appliance to consume its energy in a cost-effective way given the user preference for low delay. 

Tables \ref{table_dis_1} and \ref{table_dis_2} show that the user can control the appliance execution delay using the disutility factors for that appliances. The scenarios also show that if the user desires more comfort by minimizing appliance execution delay, the total cost may increase  because the user preferences limit the choices available to the optimizer, resulting in an appliance schedule that is more costly.  
\enlargethispage*{10pt}
\subsection{Demand Response}
The Demand Response (DR) program is designed to encourage users to change their normal energy consumption patterns in response to changes in energy price. In this section, we first explore the impact of a flat rate electricity price on the load scheduling. Then, we expand the scenario to examine the impact of Time Of Use (TOU) prices. The Ontario energy board uses TOU electricity prices which vary by time of day, day of the week (weekday or weekend), and season (winter or summer). Finally, we demonstrate the impact of dynamic prices or Real-Time Pricing (RTP) on the household load. In RTP, the energy price may change hourly and is typically announced in advance, for example a day before the real consumption takes place.
\begin{table}[!t]
\renewcommand{\arraystretch}{1.3}
\caption{Grid Energy Demand in Flat Grid Price Scheme (Household 1)}
\label{table_dr_1}
\centering
\begin{tabular}{>{\raggedright}p{.9cm} >{\raggedright}p{1.2cm} >{\centering}p{.4cm} >{\centering}p{0.4cm} >{\centering}p{0.3cm} >{\centering}p{0.3cm} >{\centering}p{0.3cm} >{\centering}p{0.3cm} >{\centering}p{0.3cm} >{\centering}p{0.3cm} }
\hline
Timeslot & {} & 1 & 2 & 3 & 4 & 5 & 6 & 7 & 8 \tabularnewline
\hline
Price	&Grid	&1	&1	&1	&1	&1	&1&	1	&1\tabularnewline
\hline
{Energy } & Grid	&20&20&	20&	20&	20&	20	&20	&20\tabularnewline
{Availa-}				& Storage	&3.06	&3.03	&3	&3.77	&3.09	&3.06	&3.03	&3\tabularnewline
{bility}													& Renewables	&0	&	0		&0		&2	&	1		&2		&0	&	0\tabularnewline
\hline
{Energy} & Grid	&1.29	&1	&1	&0	&3.36	&2	&0	&0\tabularnewline
{Source}							& Storage	&0.71	&0	&0	&0	&0.64	&0	&0	&0\tabularnewline
{}													& Renewables	&0	&	0	&	0	&	2		&1		&2	&	0		&0\tabularnewline
\hline
{Load} & Storage Charging	&1	&	0	&	0		&1	&	0	&	0		&0		&0\tabularnewline
{}											 		& App1	&1		&1		&1		&1	&	1	&	0	&	0		&0\tabularnewline
{}													& App2	&0	&0	&0	&0	&1	&1	&0	&0\tabularnewline
\hline
\multicolumn{10}{c}{Energy Cost = 8.65, Disutility Cost = 0.04, and Total Cost = 8.69} \tabularnewline
\hline
\end{tabular}
\end{table}
\begin{table}[!t]
\renewcommand{\arraystretch}{1.3}
\caption{Grid Energy Demand in TOU Price Scheme (Household 1)}
\label{table_dr_2}
\centering
\begin{tabular}{>{\raggedright}p{.9cm} >{\raggedright}p{1.2cm} >{\centering}p{.4cm} >{\centering}p{0.4cm} >{\centering}p{0.3cm} >{\centering}p{0.3cm} >{\centering}p{0.3cm} >{\centering}p{0.3cm} >{\centering}p{0.3cm} >{\centering}p{0.3cm} }
\hline
Timeslot & {} & 1 & 2 & 3 & 4 & 5 & 6 & 7 & 8 \tabularnewline
\hline
Price	&Grid	&1	&1	&2	&2	&3 &2&	1	&1\tabularnewline
\hline
{Energy } & Grid	&20&20&	20&	20&	20&	20	&20	&20\tabularnewline
{Availa-}				& Storage	&3.06	&3.03	&3	&3.77	&3.73	&4.49	&3.45	&3\tabularnewline
{bility}													& Renewables	&0	&	0		&0		&2	&	1		&2		&0	&	0\tabularnewline
\hline
{Energy} & Grid	&4.29	&4	&0	&0	&0	&0	&0	&0.58\tabularnewline
{Source}							& Storage	&0.71	&0	&0	&0	&0	&0	&1	&0.42\tabularnewline
{}													& Renewables	&0	&	0	&	0	&	2		&1		&2	&	0		&0\tabularnewline
\hline
{Load} & Storage Charging	&1	&0	&0	&1	&0	&1	&0	&0\tabularnewline
{}											 		& App1	&0	&0	&0	&1	&1	&1	&1	&1\tabularnewline
{}													& App2	&1	&1	&0	&0	&0	&0	&0	&0\tabularnewline
\hline
\multicolumn{10}{c}{Energy Cost = 8.88, Disutility Cost = 0.03, and Total Cost = 8.91} \tabularnewline
\hline
\end{tabular}
\end{table}
\begin{table}[!t]
\renewcommand{\arraystretch}{1.3}
\caption{Grid Energy Demand in RTP Scheme (Household 1)}
\label{table_dr_3}
\centering
\begin{tabular}{>{\raggedright}p{.9cm} >{\raggedright}p{1.2cm} >{\centering}p{.4cm} >{\centering}p{0.4cm} >{\centering}p{0.3cm} >{\centering}p{0.3cm} >{\centering}p{0.3cm} >{\centering}p{0.3cm} >{\centering}p{0.3cm} >{\centering}p{0.3cm} }
\hline
Timeslot & {} & 1 & 2 & 3 & 4 & 5 & 6 & 7 & 8 \tabularnewline
\hline
Price	&Grid	&0.7	&1	&0.8	&1.1	&0.6	&0.7	&1.2	&0.5\tabularnewline
\hline
{Energy } & Grid	&20&20&	20&	20&	20&	20	&20	&20\tabularnewline
{Availa-}				& Storage	&3.06	&3.03	&3	&3.77	&3.73	&3.06	&3.03	&3\tabularnewline
{bility}													& Renewables	&0	&	0		&0		&2	&	1		&2		&0	&	0\tabularnewline
\hline
{Energy} & Grid	&1.29	&0	&0	&0	&4	&2.37	&0	&1\tabularnewline
{Source}							& Storage	&0.71	&0	&0	&0	&0	&0.63	&0	&0\tabularnewline
{}													& Renewables	&0	&	0	&	0	&	2		&1		&2	&	0		&0\tabularnewline
\hline
{Load} & Storage Charging	&1	&0	&0	&1	&0	&0	&0	&0\tabularnewline
{}											 		& App1	&1	&0	&0	&1	&1	&1	&0	&1\tabularnewline
{}													& App2	&0	&0	&0	&0	&1	&1	&0	&0\tabularnewline
\hline
\multicolumn{10}{c}{Energy Cost = 5.46, Disutility Cost = 0.07, and Total Cost = 5.53} \tabularnewline
\hline
\end{tabular}
\end{table}
Tables \ref{table_dr_1}, \ref{table_dr_2} and \ref{table_dr_3} summarize the impact of these three different energy price schemes, using the same format as before. Table \ref{table_dr_1} shows the energy drawn from the grid spread over the time horizon when the grid energy price is flat. The utility has no control over the energy consumption of the user. In this case, the delay of the appliance operation is only controlled by the disutility factor. Table \ref{table_dr_2} shows that the scheduler schedules appliances in a way that no energy needs to be drawn from the grid during higher-price periods when the TOU scheme is used. For this scenario, timeslots 1, 2, 7 and 8 are off-peak; timeslot 3, 4 and 6 are mid-peak and timeslot 5 is on-peak. Results show that the household did not buy energy from the grid in mid-peak and on-peak time periods. Table \ref{table_dr_3} shows how a RTP scheme increases the grid energy demand in lower price timeslots. The grid price is comparatively low in timeslots 1, 5 and 8. The optimizer draws energy from the grid during these timeslots.
\enlargethispage*{10pt}\subsection{Pareto Optimality }
If we solve the Objective Function (\ref{eqn_25}) with Constraints  (\ref{eqn_5}) - (\ref{eqn_22}), excluding Constraint (\ref{eqn_26}), it may distribute the cost unfairly between the households. This section describes this problem in more detail and how the proposed approach solves this problem by ensuring Pareto optimality.  
\subsubsection{Cost without Microgrid Trading}
Tables \ref{table_dis_1} and \ref{table_noT_1} show the minimum costs of household 1 and household 2 respectively when the households are not participating in microgrid trading. These costs should be the maximum costs when they participate in microgrid trading for cost optimization. The minimum costs for household 1 and household 2 are 6.99 and 7.57 respectively. Therefore, when the 2 households are optimized individually, the total cost is 14.56. 
\begin{table}[!t]
\renewcommand{\arraystretch}{1.3}
\caption{Cost for Household 2 without Microgrid Trading}
\label{table_noT_1}
\centering
\begin{tabular}{>{\raggedright}p{.9cm} >{\raggedright}p{1.2cm} >{\raggedleft}p{.4cm} >{\raggedleft}p{0.4cm} >{\raggedleft}p{0.3cm} >{\raggedleft}p{0.3cm} >{\raggedleft}p{0.3cm} >{\raggedleft}p{0.3cm} >{\raggedleft}p{0.3cm} >{\raggedleft}p{0.3cm} }
\hline
Timeslot & {} & 1 & 2 & 3 & 4 & 5 & 6 & 7 & 8 \tabularnewline
\hline
Price	&Grid	&0.7&	1	&1.2	&1.5	&2	&1.7&	1.5&	0.5\tabularnewline
\hline
{Energy } & Grid	&20&20&	20&	20&	20&	20	&20	&20\tabularnewline
{Availa-}				& Storage	&4.95	&4.9	&4.85	&4.8	&3.75	&3.06	&3.03	&3\tabularnewline
{bility}													& Renewables	&1	&2	&0	&0	&1	&1	&0	&2\tabularnewline
\hline
{Energy} & Grid	&2	 &1	&0	&0	&0	&0.34	&2	&3\tabularnewline
{Source}							& Storage	&0	&0	&0	&0	&1	&0.66	&0	&0\tabularnewline
{}													& Renewables	&1	&2	&0	&0	&1	&1	&0	&2\tabularnewline
\hline
{Load} & Storage Charging	&0	&0	&0	&0	&0	&0	&0	&0\tabularnewline
{}											 		& App1	&1	&1	&0	&0	&0	&0&0	&1\tabularnewline
{}													& App2	&0	&0	&0	&0	&1	&1	&1	&1\tabularnewline
\hline
\multicolumn{10}{c}{Energy Cost = 7.48, Disutility Cost = 0.09, and Total Cost = 7.57} \tabularnewline
\hline
\end{tabular}
\end{table}

\subsubsection{Cost with Microgrid Trading}
Now, we introduce the microgrid and solve the same optimization problem, but this time also allowing households to trade energy with each other. Tables \ref{table_noPO_1} and \ref{table_noPO_2} summarize the results for the two households, using the same format as before. We added one line to the price section, as we now also keep track of the prices in the microgrid. The energy traded through the microgrid is shown in the used energy section: a negative value indicates that the household is actually selling energy into the microgrid, whereas a positive value indicates that the household buys that much energy from the microgrid. Compared to the previous solution, summarized in Table \ref{table_dis_1}, the total cost of household 1 increases from 6.99 to 12.65.
\begin{table*}[!t]
\renewcommand{\arraystretch}{1.3}
\caption{Cost for Household 1 with Microgrid Trading}
\label{table_noPO_1}
\centering
\begin{tabular}{>{\raggedright}p{1.2cm} >{\raggedright}p{2.5cm} >{\centering}p{.6cm} >{\centering}p{0.6cm} >{\centering}p{0.6cm} >{\centering}p{0.6cm} >{\centering}p{0.6cm} >{\centering}p{0.6cm} >{\centering}p{0.6cm} >{\centering}p{0.6cm} }
\hline
Timeslot & {} & 1 & 2 & 3 & 4 & 5 & 6 & 7 & 8 \tabularnewline
\hline
Price	&Grid	&0.7&	1	&1.2	&1.5	&2	&1.7&	1.5&	0.5\tabularnewline
{} &Microgrid	&0	&0	&0	&0	&0	&0	&0	&0\tabularnewline
\hline
{Energy } & Grid	&20&20&	20&	20&	20&	20	&20	&20\tabularnewline
{Availability}				& Storage	&3.77	&3.73	&3.57	&3.54	&3.5	&3.47	&3.03	&3\tabularnewline
{}													& Renewables	&0	&0	&0	&2	&1	&2	&0	&0\tabularnewline
\hline
{Microgrid} &{(Demand/Availability)}	&-2.77	&-0.75	&0.31	&-2.54	&-1.5	&-1.47	&-0.43	&-3\tabularnewline
\hline
{Energy} & Grid	&8	&5.02	&0	&0	&0	&0	&0	&4\tabularnewline
{Source}							& Storage	&0&0	&0.12	&0&0	&0	&0.40	&0\tabularnewline
{}													& Renewables	&0 &0	&0	&2	&1	&2	&0	&0\tabularnewline
\hline
{Microgrid} &{(Source/Load)}	&-2	&-0.02	&0.88	&-2	&-1	&-1	&-0.4	&-3\tabularnewline
\hline
{Load} & Storage Charging	&1	&0	&0	&0	&0	&0	&0	&0\tabularnewline
{}											 		& App1	&1	&1	&1	&0	&0	&1&0	&1\tabularnewline
{}													& App2	&1&1	&0	&0	&0	&0	&0	&0\tabularnewline
\hline
\multicolumn{10}{c}{Energy Cost = 12.62, Disutility Cost = 0.03, and Total Cost = 12.65} \tabularnewline
\hline
\end{tabular}
\end{table*}
\begin{table*}[!t]
\renewcommand{\arraystretch}{1.3}
\caption{Cost for Household 2 with Microgrid Trading}
\label{table_noPO_2}
\centering
\begin{tabular}{>{\raggedright}p{1.2cm} >{\raggedright}p{2.5cm} >{\centering}p{.6cm} >{\centering}p{0.6cm} >{\centering}p{0.6cm} >{\centering}p{0.6cm} >{\centering}p{0.6cm} >{\centering}p{0.6cm} >{\centering}p{0.6cm} >{\centering}p{0.6cm} }
\hline
Timeslot & {} & 1 & 2 & 3 & 4 & 5 & 6 & 7 & 8 \tabularnewline
\hline
Price	&Grid	&0.7&	1	&1.2	&1.5	&2	&1.7&	1.5&	0.5\tabularnewline
{} &Microgrid	&0	&0	&0	&0	&0	&0	&0	&0\tabularnewline
\hline
{Energy } & Grid	&20&20&	20&	20&	20&	20	&20	&20\tabularnewline
{Availability}				& Storage	&4.95	&3.92	&3	&4.77	&4.72	&4.68	&3.03	&3\tabularnewline
{}													& Renewables	&1	&2	&0	&0	&1	&1	&0	&2\tabularnewline
\hline
{Microgrid} &{(Demand/Availability)}	&0.05	&-0.9	&-0.88	&0.23	&-0.72	&-0.68	&0.37	&3\tabularnewline
\hline
{Energy} & Grid	&0	&0	&0	&0	&0	&0	&0	&0\tabularnewline
{Source}							& Storage	&0	&0.98	&0.88	&0	&0	&0	&1.6	&0\tabularnewline
{}													& Renewables	&1	&2	&0	&0	&1	&1	&0	&2\tabularnewline
\hline
{Microgrid} &{(Source/Load)}	&2	&0.02&-0.88	&2	&1	&1	&0.4	&3\tabularnewline
\hline
{Load} & Storage Charging	&0	&0	&0	&1	&0	&0	&0	&0\tabularnewline
{}											 		& App1	&1	&1	&0	&0	&0	&0&0	&1\tabularnewline
{}													& App2	&0&0	&0	&0	&1	&1	&1	&1\tabularnewline
\hline
\multicolumn{10}{c}{Energy Cost = 0, Disutility Cost = 0.09, and Total Cost = 0.09} \tabularnewline
\hline
\end{tabular}
\end{table*}
As Tables \ref{table_noPO_1} and \ref{table_noPO_2} show, the two households balance the energy flow in the microgrid: whenever household 1 draws on energy from the microgrid, household 2 provides that amount into the microgrid and vice versa. Table \ref{table_noPO_2} shows that household 2 does not buy any energy from the grid. Household 1 buys the required energy for household 2 from the grid and sells it for free to household 2. The scenario shows that the energy cost of household 2 is 0 and it increases the cost of household 1 through microgrid energy trading. Overall, the sum of the energy costs for both households decreased from 14.56 to 12.74. But unless there is a separate mechanism to reimburse household 1 for the additional costs, we do not believe that this solution is a desirable outcome. Trading through the microgrid should improve or at least not worsen each individual household. In other words, we require a Pareto-optimal solution, where no household can improve their energy consumption costs further without worsening another household. The optimization problem as outlined above does not deal with this issue, as it only minimizes the total energy costs, irrespective of the impact on individual households.\enlargethispage*{10pt}

Therefore, to achieve Pareto-optimal costs, we used Constraint (\ref{eqn_26}) to solve the unfair cost distribution problem. The constraint defines maximum cost bounds for each household while trading in the microgrid. The maximum cost of each household is the cost that a household pays when it does not participate in microgrid energy trading. This then rules out solutions such as the one presented in Table \ref{table_noPO_1} for household 1, whose energy cost increased from 7.57 to 12.65, which would violate that upper bound. The resulting solution to the unified optimization problem will then be one where each household is no worse off than in the absence of the microgrid, and no further improvements are possible, representing a Pareto-optimal point.	

Tables \ref{table_PO_1} and \ref{table_PO_2} show the results when applying this method on the small case study. As before, trading via the microgrid is balanced, and some timeslots see microgrid prices different from 0 to clear the market (determined by the optimization problem). The energy costs for both households are lower than in the absence of the microgrid. They dropped from 6.99 to 6.87 for household 1 and from 7.57 to 5.86 for household 2. The total energy cost for both households is 12.74, which is lower than the initial total energy costs of 14.56.
\begin{table*}[!t]
\renewcommand{\arraystretch}{1.3}
\caption{Cost for Household 1 - Pareto-Optimal Cost with Microgrid}
\label{table_PO_1}
\centering
\begin{tabular}{>{\raggedright}p{1.2cm} >{\raggedright}p{2.5cm} >{\centering}p{.6cm} >{\centering}p{0.6cm} >{\centering}p{0.6cm} >{\centering}p{0.6cm} >{\centering}p{0.6cm} >{\centering}p{0.6cm} >{\centering}p{0.6cm} >{\centering}p{0.6cm} }
\hline
Timeslot & {} & 1 & 2 & 3 & 4 & 5 & 6 & 7 & 8 \tabularnewline
\hline
Price	&Grid	&0.7&	1	&1.2	&1.5	&2	&1.7&	1.5&	0.5\tabularnewline
{} &Microgrid	&0.62	&0.66	&0.17	&0	&1.05	&1.17	&0.3	&0.35\tabularnewline
\hline
{Energy } & Grid	&20&20&	20&	20&	20&	20	&20	&20\tabularnewline
{Availability}				& Storage	&3.77	&3.37	&3.34	&3.31	&3.27	&3.24	&3.03	&3\tabularnewline
{}													& Renewables	&0	&0	&0	&2	&1	&2	&0	&0\tabularnewline
\hline
{Microgrid} &{(Demand/Availability)}	&-2.77	&2.53	&0.66	&-2.31	&-1.27	&-1.24	&-0.21	&-3\tabularnewline
\hline
{Energy} & Grid	&8	&1.74	&0	&0	&0	&0	&0	&4\tabularnewline
{Source}							& Storage	&0	&0.36	&0	&0	&0	&0	&0.18	&0\tabularnewline
{}													& Renewables	&0 &0	&0	&2	&1	&2	&0	&0\tabularnewline
\hline
{Microgrid} &{(Source/Load)}	&-2	&2.9	&1	&-2	&-1	&-1	&-0.18	&-3\tabularnewline
\hline
{Load} & Storage Charging	&1	&0	&0	&0	&0	&0	&0	&0\tabularnewline
{}											 		& App1	&1	&1	&1	&0	&0	&1&0	&1\tabularnewline
{}													& App2	&1&1	&0	&0	&0	&0	&0	&0\tabularnewline
\hline
\multicolumn{10}{c}{Energy Cost = 6.84, Disutility Cost = 0.03, and Total Cost = 6.87} \tabularnewline
\hline
\end{tabular}
\end{table*}
\begin{table*}[!t]
\renewcommand{\arraystretch}{1.3}
\caption{Cost for Household 2 - Pareto-Optimal Cost with Microgrid}
\label{table_PO_2}
\centering
\begin{tabular}{>{\raggedright}p{1.2cm} >{\raggedright}p{2.5cm} >{\centering}p{.6cm} >{\centering}p{0.6cm} >{\centering}p{0.6cm} >{\centering}p{0.6cm} >{\centering}p{0.6cm} >{\centering}p{0.6cm} >{\centering}p{0.6cm} >{\centering}p{0.6cm} }
\hline
Timeslot & {} & 1 & 2 & 3 & 4 & 5 & 6 & 7 & 8 \tabularnewline
\hline
Price	&Grid	&0.7&	1	&1.2	&1.5	&2	&1.7&	1.5&	0.5\tabularnewline
{} &Microgrid	&0.62	&0.66	&0.17	&0	&1.05	&1.17	&0.3	&0.35\tabularnewline
\hline
{Energy } & Grid	&20&20&	20&	20&	20&	20	&20	&20\tabularnewline
{Availability}				& Storage	&4.95	&4.28	&3.23	&5	&4.95	&4.9	&3.03	&3\tabularnewline
{}													& Renewables	&1	&2	&0	&0	&1	&1	&0	&2\tabularnewline
\hline
{Microgrid} &{(Demand/Availability)}	&0.05	&-4.18	&-1.23	&0	&-0.95	&-0.9	&0.15	&3\tabularnewline
\hline
{Energy} & Grid	&0	&3.28	&0	&0	&0	&0	&0	&0\tabularnewline
{Source}							& Storage	&0	&0.63	&1	&0	&0	&0	&1.82	&0\tabularnewline
{}													& Renewables	&1	&2	&0	&0	&1	&1	&0	&2\tabularnewline
\hline
{Microgrid} &{(Source/Load)}	&2	&-2.9	&-1	&2	&1	&1	&0.18	&3\tabularnewline
\hline
{Load} & Storage Charging	&0	&0	&0	&1	&0	&0	&0	&0\tabularnewline
{}											 		& App1	&1	&1	&0	&0	&0	&0&0	&1\tabularnewline
{}													& App2	&0&0	&0	&0	&1	&1	&1	&1\tabularnewline
\hline
\multicolumn{10}{c}{Energy Cost = 5.77, Disutility Cost = 0.09, and Total Cost = 5.86} \tabularnewline
\hline
\end{tabular}
\end{table*}
\enlargethispage*{10pt}
\section{Complexity Analysis}
The proposed unified MINLP model could be first reduced to a MILP problem in polynomial time then the resulting MILP could be restricted to have only real numbers. Hence, the proposed model reduces to an Integer Programming (IP) problem in polynomial time. Garey and Johnson proved that IP is an NP-hard problem \cite{Michael1979}. Even if the variables are restricted to {0, 1} it remains NP-hard \cite{Michael1979}. Therefore, the proposed non-convex MINLP model is NP-hard. \enlargethispage*{10pt}

\begin{figure}[!t]
\centering
\includegraphics[width=3.5in]{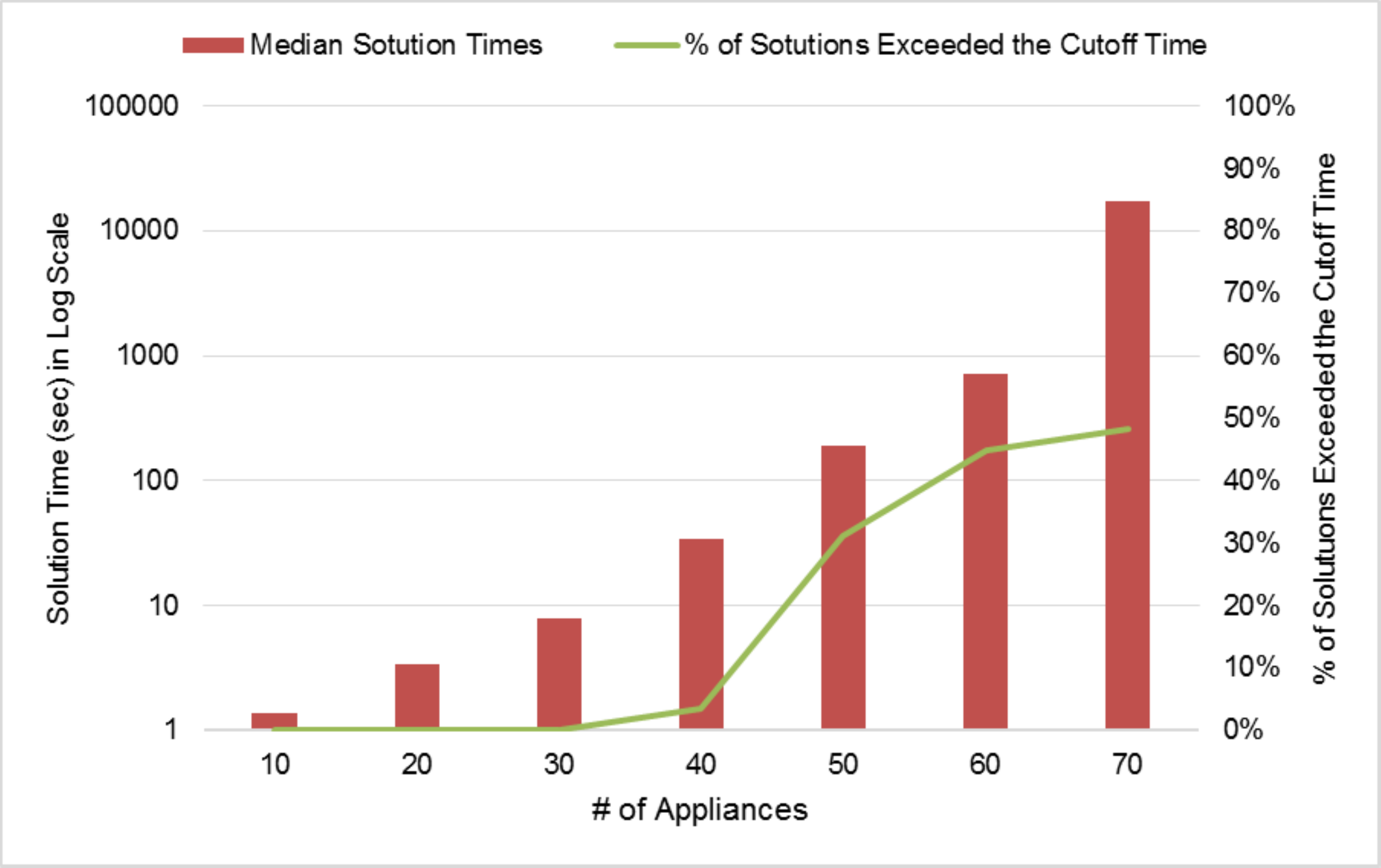}
\caption{Solution Time vs. Number of Appliances.}
\label{fig_app}
\end{figure}

We explored experimentally the impact of the number of appliances, timeslots, and households on the solution time. The simulation environment consists of static and random parameters. All households are considered to have similar storages \textit{i.e.}, with the same efficiency and self-discharging rate but different storage capacities. An appliance operation request occurs at the 1st timeslot and it can be executed until the last timeslot. The maximum power limit of a household is set to a large value (\textit{e.g.}, 200,000). Initial storage energy is 5 for both households. The static parameters are specified in Table \ref{table_sta_param}. The minimum and maximum bound for generating random parameters are presented in Table \ref{table_ran_param}. The duration of an appliance is randomly generated between 1 and the total number of timeslots. An appliance consumes energy uniformly random between 0.5 and 15 units per timeslot. The grid energy price at any timeslot various uniformly random from 0.1 to 5. The disutility factor of an appliance is randomly selected from any value between 0.01 and 10. The required storage charging power is generated from a random uniform distribution in the range of 2 to 5. The minimum capacity of a storage stays between 0 to 2 and the maximum capacity of a storage is limited between 5 to 10. The amount of renewable generation is chosen from a range of 0 to 10 per timeslot. 
\begin{table}[!t]
\renewcommand{\arraystretch}{1.3}
\caption{Static Parameters}
\label{table_sta_param}
\centering
\begin{tabular}{l>{\centering}p{3cm}}
\hline
Parameter	& Value \tabularnewline
\hline
Storage Efficiency	 & 90\% \tabularnewline
Storage Self-discharging Rate&	0.01\% \tabularnewline
Storage Initial Energy (Unit)	&5 \tabularnewline
Reservation time&	1st Timeslot \tabularnewline
End Time&	Last Timeslot \tabularnewline
Maximum Power Limit  \tabularnewline of a Household (Unit)	&Infinite (200,000) \tabularnewline
\hline
\end{tabular}
\end{table}
\begin{table}[!t]
\renewcommand{\arraystretch}{1.3}
\caption{Random Parameters}
\label{table_ran_param}
\centering
\begin{tabular}{l>{\centering}p{1.2cm}>{\centering}p{1.8cm}}
\hline
Parameter	& Min & Max \tabularnewline
\hline
Duration of Appliance Operations &	1	& Total Timeslots	\tabularnewline
Appliance Power (Unit)	&0.5	&15	\tabularnewline
Grid Energy Price (Unit)&	0.1	&5	\tabularnewline
Disutility Factor	&0.01	&10	\tabularnewline
Storage Power (Unit)	&2	&5	\tabularnewline
Minimum Storage Capacity (Unit)	&0	&2	\tabularnewline
Maximum Storage Capacity (Unit)	&5	&10	\tabularnewline
Renewable Energy (Unit)	&0	&10	\tabularnewline
\hline
\end{tabular}
\end{table}
Figure \ref{fig_app} shows the relationship between the median solution time and the number of appliances. At least 2 households are required to form a microgrid. Only 3 timeslots are considered arbitrarily. The number of timeslots and households are set to these minimum levels. For the same number of appliances, the simulations were run for 29 random instances. The bar diagram shows the median solution time and the line diagram shows the percentages of solution instances which exceeded the cut off execution time. The cut off execution time is imposed by the NEOS server. It does not allow a submitted problem to run more than 8 hours. The simulation terminates if the execution time exceeds 8 hours. Figure \ref{fig_app} shows that for the minimum number of households and timeslots, the median solution times increase exponentially when increasing the number of appliances.

\begin{figure}[!t]
\centering
\includegraphics[width=3.5in]{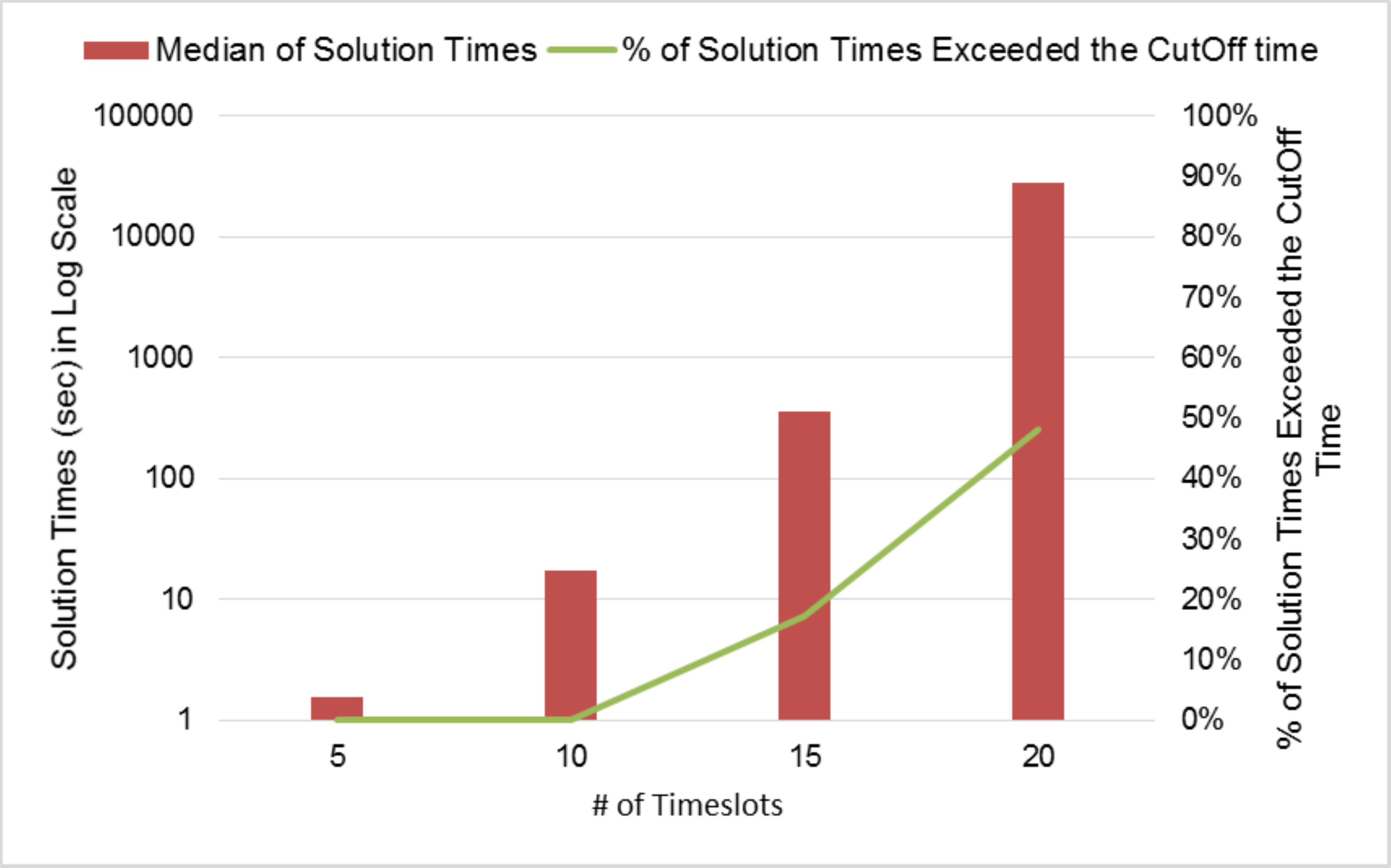}
\caption{Solution Time vs. Number of Timeslots.}
\label{fig_ts}
\end{figure}

\begin{figure}[!t]
\centering
\includegraphics[width=3.5in]{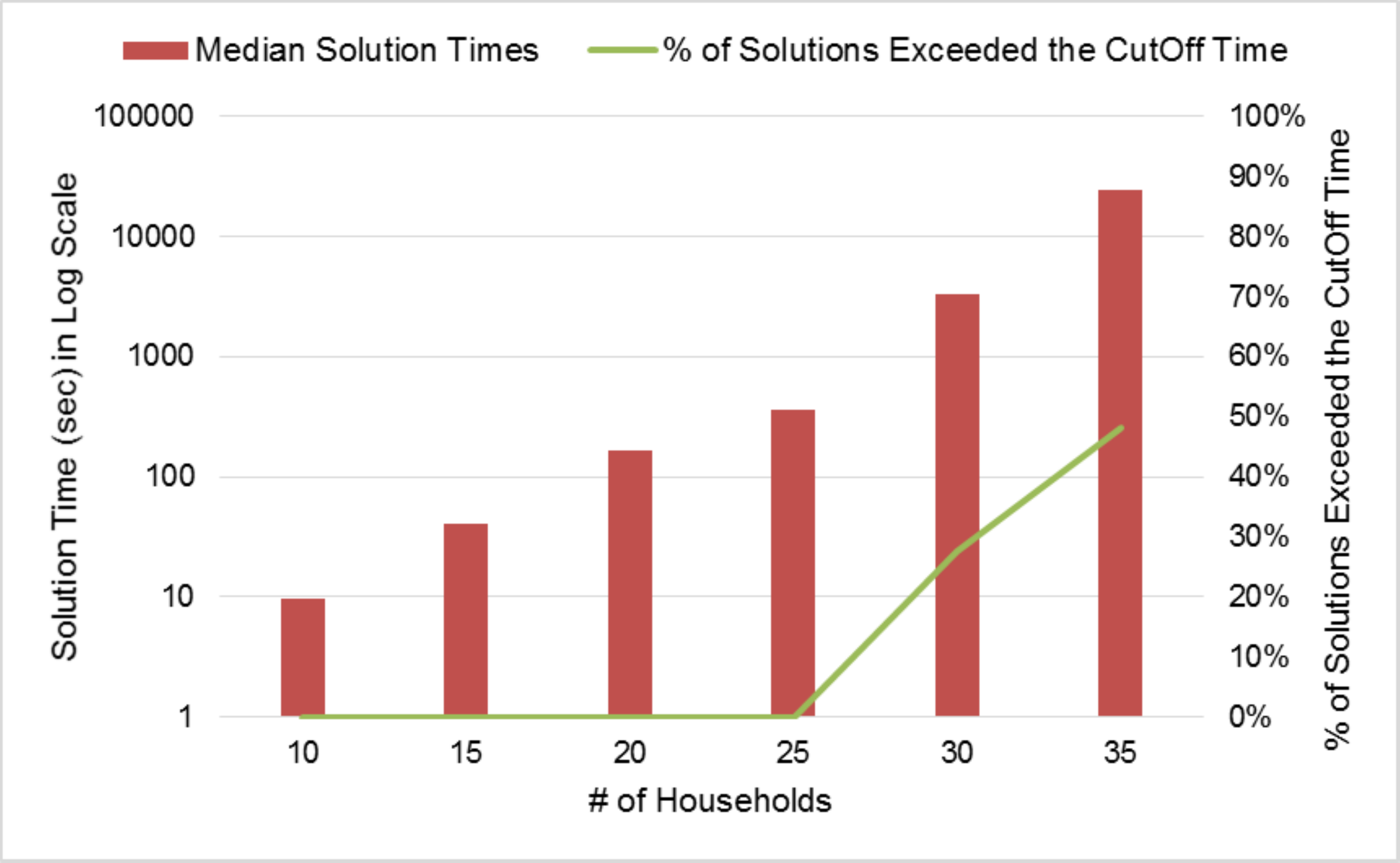}
\caption{Solution Time vs. Number of Households.}
\label{fig_house}
\end{figure}

Figure \ref{fig_ts} illustrates that the median solution time increases exponentially with the number of timeslots for the minimum number of appliances (2 appliances: 1 interruptible and 1 uninterruptible) and the minimum number of households (2 households). For the same number of timeslots, the simulations were run for 29 random instances. Finally, Figure \ref{fig_house} shows that the time complexity increases exponentially with the number of households for the minimum number of appliances and timeslots. The simulations were executed for 29 random instances for the same number of households. 

The results show that the proposed problem is NP-hard. Even for small problem instances, solution times increase exponentially.  

%
%

\section{Conclusion}
This research explores the contemporary cost saving methods for the smart grid from the users' perspective. It identified the primary strategies and components for cost optimization. The proposed non-convex MINLP model has achieved the primary objectives of the research. It does not only consider optimal energy cost optimization but also addresses the inconvenience created to the users by delaying certain tasks. The constraints, variables and parameters define the underlying features of the smart grid components, \textit{e.g.}, load characteristics, source behavior and user preferences. The model determines the energy price in the energy market and regulates the demand and supply of microgrid energy. It minimizes the energy buying cost from the utility in a microgrid area as well as ensures Pareto optimality among the participating households while trading energy. The proposed model unified the partial models identified in previous research in a single unified optimization model to attain optimal cost saving profiles for the participating smart homes.

The proposed model is an NP-hard problem which means that using the resulting solution is not practical because the time complexity increases exponentially according to the increase in the problem size. The numerical results illustrate that even for small problem sizes, the solution time grows exponentially. In the smart grid, the price signals, the energy demand and the renewable energy generation change frequently within an hour demanding to re-solve the optimization problem with updated information. Therefore, the optimization problem should be solved in a realistic timeframe so that updates can be reflected back to the system to provide an optimized cost. Hence, aiming for an optimal solution may not be a computationally efficient approach for realistic scenarios. We need to trade off the optimal solution for reduced computation time. 

The proposed model considers the smart home as the building unit of the smart grid. This concept is also applicable to the apartments of a smart building. Therefore, the proposed model can be applied to analyze the behavior of a population (\textit{e.g.}, smart buildings, smart cities, communities, metropolitans, etc.) and properties of the components for specific scenarios. As an example, the model considers the storage as a component of smart homes. Recent advances of EV technologies imply that EVs will become an integral part of smart cities in the future. The EVs can be represented as storage to analyze the impact of EVs on the smart grid. Similarly, the proposed model can also be used to analyze the impact of renewable energy. It could be utilized to determine the quantity of resources to build a net-zero energy community or building or city. Hence, the proposed model could be useful for governments, policy makers and the utilities to predict the users' behavior and conduct a cost-benefit analysis.


%





\ifCLASSOPTIONcaptionsoff
  \newpage
\fi



\bibliographystyle{IEEEtran}
\bibliography{IEEEabrv,Unified_Paper_References}
\end{document}